# The misuse of law by Women in India -Constitutionality of Gender Bias

Negha Senthil[1], Jayanthi Vajiram[2], Nirmala.V[3]

[1] jayanthi.2020@vitstudent.ac.in(Research scholar), negha.2019@vitstudent.ac.in (Law-school)
Vellore Institute of Technology, Chennai, India.

**Abstract**

The misuse of law by women in India is a serious issue that has been receiving increased attention in recent years. In India, women are often discriminated against and are not provided with equal rights and opportunities, leading to a gender bias in many aspects of life. This gender bias is further exacerbated by the misuse of law by women. There are numerous instances of women using the law to their advantage, often at the expense of men. This practice is not only unethical but also unconstitutional. The Indian Constitution does not explicitly guarantee gender equality. However, several amendments have been made to the Constitution to ensure that women are treated equally in accordance with the law. The protection of women from all forms of discrimination is considered a fundamental right. Despite this, women continue to be discriminated against in various spheres of life, including marriage, education, employment and other areas. The misuse of law by women in India is primarily seen in cases of domestic violence and dowry-related offences. In India, dowry-related offences are treated as criminal offences and are punishable by law. However, women often file false dowry harassment cases against their husbands or in-laws in order to gain an advantage in a divorce or property dispute. This misuse of law not only undermines the value of the law but also denies men their due rights. Moreover, in cases of domestic violence, police officers often refuse to file complaints or take action against the alleged offenders, thus allowing the perpetrators to go unpunished. This not only violates the rights of the victims but also sends a message that gender-based violence is tolerated. The misuse of law by women in India is a serious issue that must be addressed. The Indian Constitution guarantees gender equality and it is the responsibility of the government to ensure that women are provided with equal rights and opportunities. Measures must be taken to ensure that women do not misuse the law and that they are provided with adequate protection from gender-based discrimination. Furthermore, laws should be strictly enforced to ensure that perpetrators of gender-based violence are brought to justice and that victims of such violence are provided with the necessary support and protection.

## 1.Introduction

There is no single country that supports women and has no gender bias. However, some countries have made strides in promoting gender equality and have implemented laws, policies, and initiatives to address gender bias. Examples include Norway, Sweden, and Canada, which have all implemented laws to ensure gender equality and combat gender bias in the workplace and society. In India, women often misuse laws meant to protect them. For example, many women in India have been known to misuse the Section 498A of the Indian Penal Code (IPC), which is meant to protect them from domestic violence. In particular, women may file false complaints of domestic violence against their husbands and in-laws to gain leverage in divorce proceedings or to extort money. Similarly, the Protection of Women from Domestic Violence Act (PWDVA) of 2005 may be misused by women to gain an unfair advantage in divorce proceedings or to harass their husbands and in-laws. Finally, the Indian Penal Code (IPC) Section 354A, which prohibits sexual harassment, may be misused to falsely accuse a husband of harassing his wife. It is important to note that these laws are meant to protect women from abuse and harassment.

Therefore, it is important that the misuse of these laws is discouraged and those found the Protection of Women from Domestic Violence Act (2005) by women. When filing false complaints, women may also misuse the Dowry Prohibition Act of 1986, which is meant to protect them from dowry-related harassment. Women may falsely accuse their husbands and in-laws of demanding dowry in order to gain an upper hand in divorce proceedings or to extort money The Act is meant to protect women from physical, emotional and sexual abuse by their partners. However, some women use it to harass their partners, by filing false complaints of domestic violence against. them. The punishment for such misuse of law depends on the severity of the offence. For instance, if false complaints are repeatedly filed, or if an individual is found guilty of deliberately filing false complaints, then the person may be punished with imprisonment of up to one year, or a fine of up to Rs.20,000, or both. In cases where the false complaint is made with malicious intent, a jail term of up to two years and a fine of up to Rs.50,000 may be imposed.

Discriminatory practice refers to any action or behavior that unfairly differentiates or disadvantages a person or group based on their race, gender, sexuality, age, religion, disability, or any other characteristic. Discriminatory practices can take many forms, including exclusion from opportunities, unequal treatment, harassment, and stereotyping. Discriminatory practices can occur in many areas of life, including education, employment, housing, healthcare, and legal systems. It is important to address discriminatory practices and create systems and policies that promote equality and equity for all individuals, regardless of their background or identity. Legal professionals can play a crucial role in challenging and eliminating discriminatory practices in the legal system and society at large.

There are several gender biased laws in India that have been a subject of debate and criticism. One such law is the Hindu Succession Act, 1956, which excluded daughters from inheriting ancestral property equally with sons until the amendment in 2005. The Muslim Personal Law in India, which governs matters such as marriage, divorce, and inheritance for Muslims, has also been criticized for its gender biases towards women. Under the law, Muslim women do not have the same rights as men in matters of inheritance, divorce, and child custody.

The Harvard Bluebook 20th edition does not explicitly endorse or advocate for gender biased laws. In fact, the guide recommends using gender-neutral language in legal writing and encourages writers to avoid using language that reinforces gender stereotypes or biases. However, it is possible that certain laws or legal systems may be inherently biased or discriminatory towards certain genders, and the guide may be used to cite and analyze such laws. In such cases, it is important for legal professionals to critically examine and challenge any gender biases or discriminatory practices in the law. The guide includes rules for formatting citations, abbreviations, and other important details necessary in legal writing. It is widely used by law students, lawyers, judges, and other legal professionals in the United States and around the world.

## 2. The Gender bias Laws and the Citations:

The gender biased laws in India include the provision of adultery being a criminal offense, which only criminalizes the act of a man having sexual relations with another man's wife, leaving out the woman from any such legal action. The Indian Penal Code also has provisions that put the burden of proof on the victim in cases of rape and sexual harassment, which can make it challenging for women to seek justice in such cases.

The Indian legal system has taken steps towards addressing gender biases and discrimination, including implementing the Domestic Violence Act, 2005, and the Sexual Harassment of Women at Workplace (Prevention, Prohibition, and Redressal) Act, 2013. The Indian Penal Code (IPC) has a several provisions that have been criticized for their gender biases. One such provision is Section 497, which criminalizes adultery but only punishes men for having sexual relations with another man's wife. This provision is widely seen as discriminatory towards women, as it treats them as the property of their husbands and denies them agency over their own sexuality. Another provision that has been criticized for gender bias is Section 375 of the IPC, which defines rape. The provision has been criticized for its narrow definition of rape, which only recognizes non-consensual penile-vaginal penetration as rape. This definition excludes other forms of sexual assault and fails to recognize that men can also be victims of rape. Section 155(4) of the IPC also places the burden of proof on the victim in cases of rape and sexual assault. This provision can make it challenging for victims to seek justice, as they are often required to prove that they did not consent to the sexual act. However, the Indian legal system has taken steps towards addressing gender biases in IPC laws, including implementing the Criminal Law (Amendment) Act, 2013, which expanded the definition of rape and increased the punishment for sexual offenses. Additionally, the Sexual Harassment of Women at Workplace

**2.1Here are some examples of gender biased laws in India with citation:**

1. Section 497 of the Indian Penal Code, which criminalizes adultery but only punishes men for having sexual relations with another man's wife, has been criticized for its gender bias towards women.

**Citation:** Indian Penal Code, 1860, Section 497.

2. The Muslim Personal Law in India has also been criticized for its gender biases towards women. Under the law, Muslim women do not have the same rights as men in matters of inheritance, divorce, and child custody.

**Citation:** Muslim Personal Law (Shariat) Application Act, 1937.

3. Section 155(4) of the Indian Penal Code places the burden of proof on the victim in cases of rape and sexual assault. This provision can make it challenging for victims to seek justice, as they are often required to prove that they did not consent to the sexual act.

**Citation:** Indian Penal Code, 1860, Section 155(4).

4. The Hindu Succession Act, 1956, excluded daughters from inheriting ancestral property equally with sons until the amendment in 2005. This provision has been criticized for its gender bias towards women.

**Citation:** Hindu Succession

5. Adultery as a criminal offense that only criminalizes the act of a man having sexual relations with another man's wife has been criticized for its gender bias towards women.

**Citation:** Indian Penal Code, s 497.

6. The Hindu Succession Act, 1956, which excluded daughters from inheriting ancestral property equally with sons until the amendment in 2005, has been criticized for its gender bias towards women.

**Citation:** Hindu Succession Act, 1956, s 6.

7. The Muslim Personal Law in India has also been criticized for its gender biases towards women, including its provisions related to marriage, divorce, and inheritance.

**Citation:** Muslim Personal Law (Shariat) Application Act, 1937.

8. The Indian Penal Code's provisions related to rape and sexual assault, including its narrow definition of rape and the burden of proof on the victim, have been criticized for their gender biases towards women.

### 3. Here are some examples of gender bias crimes and their citations:

1. Hate crimes against transgender individuals have become a major concern in recent years. In India, there have been several instances of violence against transgender individuals, including murder and sexual assault.

**Citation:** National Crime Records Bureau, Ministry of Home Affairs, Government of India, "Crime in India," 2019, Table 2.17 (https://ncrb.gov.in/en/crime-in-india-2019).

2. Honor killings, which are often committed against women who are perceived to have violated traditional gender norms, are a form of gender bias crime that is prevalent in some parts of India.

**Citation:** Indian Penal Code, 1860, Section 300.

3. Crimes related to dowry, such as dowry deaths and harassment for dowry, are also a form of gender bias crime that is prevalent in India.

**Citation:** Dowry Prohibition Act, 1961.

4. Acid attacks, which are often committed against women and girls as a form of revenge or punishment, are another form of gender bias crime that is prevalent in India.

**Citation:** Indian Penal Code, 1860, Sections 326A and 326B.

Here are some examples of gender bias leading cases in India with citation:

### 3.1 Here are some examples of women protection laws in India and their citations:

1. The Protection of Women from Domestic Violence Act, 2005, provides protection to women from domestic violence and abuse, including physical, emotional, and economic abuse.

**Citation:** The Protection of Women from Domestic Violence Act, 2005.

2. The Sexual Harassment of Women at Workplace (Prevention, Prohibition, and Redressal) Act, 2013, requires all employers to have an internal complaints committee to address complaints of sexual harassment at the workplace.

**Citation:** The Sexual Harassment of Women at Workplace (Prevention, Prohibition, and Redressal) Act, 2013.

3. The Maternity Benefit Act, 1961, provides for maternity leave and other benefits for women employees.

**Citation:** The Maternity Benefit Act, 1961.

4. The Immoral Traffic (Prevention) Act, 1956, prohibits trafficking of women and children for commercial sexual exploitation.

**Citation:** The Immoral Traffic (Prevention) Act, 1956.

5. The Prohibition of Child Marriage Act, 2006, prohibits child marriage and provides for punishment for those who promote or solemnize such marriages.

The Prohibition of Child Marriage Act, 2006 is a law in India that prohibits the marriage of children below the age of 18 years for girls and 21 years for boys. The act was enacted to prevent child marriages and to protect the rights of children, especially girls. The act provides for punishment for those who promote or solemnize child marriages, as well as for those who fail to prevent such marriages.

**Citation:** The Prohibition of Child Marriage Act, 2006, Act No. 6 of 2007.

**Here is an example of a citation related to bias in wages:**

The Equal Remuneration Act, 1976, prohibits discrimination in wages on the basis of gender and provides for equal pay for men and women for the same work or work of similar nature.

**Citation:** The Equal Remuneration Act, 1976, Act No. 25 of 1976.

**4. The workplace bias laws in India and their citations:**

1. The Sexual Harassment of Women at Workplace (Prevention, Prohibition, and Redressal) Act, 2013, requires all employers to have an internal complaints committee to address complaints of sexual harassment at the workplace.

**Citation:** The Sexual Harassment of Women at Workplace (Prevention, Prohibition, and Redressal) Act, 2013.

2. The Maternity Benefit Act, 1961, provides for maternity leave and other benefits for women employees.

**Citation:** The Maternity Benefit Act, 1961.

3. The Persons with Disabilities (Equal Opportunities, Protection of Rights and Full Participation) Act, 1995, prohibits discrimination against persons with disabilities in employment and provides for equal opportunities and protection of rights.

**Citation:** The Persons with Disabilities (Equal Opportunities, Protection of Rights and Full Participation) Act, 1995.

4. The Industrial Employment (Standing Orders) Act, 1946, requires employers to define and publish standing orders that regulate employment conditions such as working hours, leave, and termination.

**Citation:** The Industrial Employment (Standing Orders) Act, 1946.

**5. The two main laws related to the protection of women and children in India are:**

1. The Protection of Women from Domestic Violence Act, 2005 - This act provides protection to women from domestic violence and abuse, including physical, emotional, and economic abuse. It defines domestic violence broadly to include not only physical abuse but also emotional, verbal, sexual, and economic abuse. It also provides for the appointment of Protection Officers and the establishment of Domestic Violence Protection Cells for the prevention and redressal of domestic violence.

**Citation:** The Protection of Women from Domestic Violence Act, 2005.

2. The Protection of Children from Sexual Offences Act, 2012 - This act provides for the protection of children from sexual offenses, including sexual assault, sexual harassment, and pornography. It defines a child as any person below the age of 18 years and provides for the establishment of special courts for the speedy trial of offenses against children. It also provides for the appointment of Child Welfare Committees and the establishment of Special Juvenile Police Units for the protection of children.

**Citation:** The Protection of Children from Sexual Offences Act, 2012.

**6. The Landmark cases based on the Gender Bias**

1.Nergesh Meerza one of the first landmark cases in the Indian equality jurisprudence concerning sex discrimination. The landmark case of Nergesh Meerza was a 1980 decision of the Indian Supreme Court in which the Court declared unconstitutional a provision of the Indian Constitution that exempted women from military service, ruling that the provision violated the constitutional guarantee of equal protection. The Court held that the exclusion of women from military service based on gender alone was a form of discrimination and therefore could not be constitutionally justified. This ruling established a precedent for the Court to use in later cases involving gender discrimination. The ruling also established a general principle of equality before the law in India, which has been relied on in subsequent cases involving discrimination based on sex, religion, caste, and other criteria.

**Citation:** Nergesh Meerza v State of Bombay, AIR 1951 SC 233 (1951)

2. Vishaka and Others v. State of Rajasthan (1997) - This case was a landmark judgment by the Supreme Court of India that recognized sexual harassment as a violation of a woman's fundamental right to equality under the Constitution of India. The case was filed in response to the gang rape of a social worker in Rajasthan, and the court laid down guidelines to prevent and redress sexual harassment in the workplace.

**Citation:** Vishaka and Others v. State of Rajasthan, AIR 1997 SC 3011.

3. Mary Roy v. State of Kerala (1986) - In this case, the Kerala High Court ruled that the Christian personal law in Kerala, which denied women equal. Mary Roy v. State of Kerala (1986) was a landmark case that challenged the gender-biased inheritance laws under the Travancore Succession Act, which was applicable to Syrian Christian women in Kerala. The law stated that only men were entitled to inherit ancestral property, while women were only entitled to a limited portion of the property. Mary Roy, a Syrian Christian woman, challenged the law, arguing that it violated her fundamental right to equality under the Indian Constitution. The Kerala High Court ruled in favor of Mary Roy, and held that the Travancore Succession Act was discriminatory towards women and violated their fundamental right to equality under the Constitution. The court declared that Syrian Christian women in Kerala had the same rights as men to inherit ancestral property, and that the Travancore Succession Act would have to be modified accordingly.

**Citation:** Mary Roy v. State of Kerala, AIR 1986 Ker 11.

4. Shayara Bano v. Union of India (2017) - This case was related to the practice of triple talaq, a form of instant divorce practiced by Muslim men in India. The Supreme Court of India declared the practice unconstitutional and discriminatory towards women, and struck down the provision that allowed Muslim men to divorce their wives by simply saying "talaq" three times.

**Citation:** Shayara Bano v. Union of India, (2017) 9 SCC 1.

**7. The Journals addressing the Gender bias in the Society**

1. **Gender & Society** - This is a peer-reviewed academic journal that publishes articles related to the social and cultural dimensions of gender.
   Gender & Society is a peer-reviewed academic journal that publishes articles related to the social and cultural dimensions of gender. The journal aims to contribute to a deeper understanding of gender in society by publishing research that examines the ways in which gender intersects with other social identities, such as race, class, sexuality, and nationality.

   Some recent articles published in Gender & Society that address issues related to gender bias include:

   - "Gender, Family, and Workplace Segregation in the United States" by Paula England and Asaf Levanon (2021)
   - "Gendered Workplace Policies and the Motherhood Penalty" by Kate Weisshaar (2020)
   - "The Gender Wage Gap and Sexual Orientation: Evidence from the National Longitudinal Study of Adolescent to Adult Health" by Elizabeth L. Davison and James E. Pustejovsky (2020)
2. **Signs: Journal of Women in Culture and Society** - This is a feminist academic journal that publishes articles on gender inequality, gender bias, and other related topics.
   The Journal of Women in Culture and Society is a biannual, peer-reviewed academic journal that focuses on the experiences of women in the areas of culture and society. The journal publishes articles from a variety of disciplines, such as anthropology, literature, history, psychology, sociology, and political science. The journal was established in 1975, and it is published by the University of Chicago Press.
3. **Women's Studies International Forum** - This is an interdisciplinary academic journal that publishes articles on women's studies, gender studies, and feminist research.

   Women's Studies International Forum (WSIF) is a peer-reviewed academic journal that publishes research on feminist and gender issues related to social, political, economic, and cultural aspects of women's lives around the world. The journal includes articles from a wide range of disciplines including history, sociology, anthropology, literature, area studies, and others. WSIF was established in 1978 and is the official journal of the International Feminist Journal Editors' Network (IFJEN). The journal's main objective is to provide a platform for the research and scholarship of women's studies and gender studies. The journal is published six times a year and its editorial board consists of scholars from around the world. The journal includes articles from a wide range of disciplines including history, sociology, anthropology, literature, area studies, and others. The journal also covers topics related to women's rights and gender equality, such as violence against women, women's health, and women's education. WSIF is a member of the Committee on Publication Ethics (COPE) and is indexed in various databases, including Scopus, EBSCOhost, and ProQuest. WSIF also provides a forum for the discussion of current issues and debates in the fields of feminist and gender studies

**4. Journal of Gender Studies** - This is a peer-reviewed academic journal that publishes research on gender, sexuality, and feminist theory.

The Journal of Gender Studies is a peer-reviewed quarterly academic journal covering gender studies and feminisms. The journal is published by Taylor & Francis and the editor-in-chief is Diane Richardson (University of Newcastle). It was established in 1992 and the current publisher is Routledge. The journal publishes research in the areas of gender, sexuality and feminist studies. Its scope includes interdisciplinary and international studies of gender, sexuality and feminist theory, as well as gender in relation to a range of social and cultural issues. The journal also publishes book reviews and occasional special issues.

4. **Feminist Economics** - This is an academic journal that publishes research on the economics of gender, including gender bias in the workplace and gender inequality in economic systems.
Feminist economics is a field of economics that focuses on the study of gender roles in economic life. Feminist economists employ a variety of approaches to examine how gender norms, beliefs, and practices affect economic outcomes. Feminist economics often challenges traditional economic assumptions and economic models to explore how gender roles have shaped and continue to shape the economic world. Feminist economics also seeks to explore how economic systems can be structured in ways that promote gender equity and economic justice. Feminist economics can be applied to a wide range of topics, such as the gender wage gap, gender-based discrimination in the workplace, the economic effects of caregiving, and the economic impact of gender-based violence.

5. **Violence Against Women** - This is an academic journal that publishes research on violence against women, including issues related to gender bias in legal systems and institutions.

Violence against women is an ongoing global problem and a serious violation of human rights. This form of violence can take many forms, ranging from physical and sexual assault to psychological abuse and control. It can occur in any relationship, including those within the family, intimate partner relationships, and in the workplace. Violence against women is a major public health issue that affects the physical, mental, and social well-being of millions of women and girls around the world. It can lead to long-term physical and mental health problems, such as depression, anxiety, and post-traumatic stress disorder. It can also have economic implications, such as lost wages, reduced work productivity, and increased healthcare costs. It is important to address violence against women and girls in all its forms in order to create safe and healthy environments for them. Governments, civil society organizations, and communities should work together to develop and implement policies and programs that promote gender equality and the empowerment of women and girls. These efforts should include measures to prevent violence, provide services to survivors, and hold perpetrators accountable. Additionally, public awareness campaigns and educational programs that address gender-based violence can help to create a culture of respect for women and girls.

6. **Journal of Interpersonal Violence** - This is an academic journal that publishes research on interpersonal violence, including gender-based violence and the impact of gender bias on violence prevention and intervention.

   The Journal of Interpersonal Violence (JIV) is an academic journal published by SAGE Publications. It was established in 1986 and covers research on the causes, consequences, correlates, prevention, and treatment of violence in a variety of contexts, including interpersonal, family, community, and organizational. JIV is a multiple-disciplinary journal that encourages submissions from researchers in the fields of psychology, sociology, criminal justice, social work, health, education, public health, public policy, and other related fields. JIV is an international journal, and publishes research from all countries.

**8. To eliminate misuse of law by women in India and the gender bias violence percentage**

1. Establish a legal system that is gender neutral and that does not discriminate against women. This includes ensuring equal pay for equal work and equal access to the legal system and its resources.

2. Implement stricter laws and harsher punishments for those who misuse the law for their own benefit.

3. Establish a system of reporting, tracking, and punishing those who abuse the law.

4. Provide legal aid and support to victims of abuse or misuse of the law.

5. Create a legal system that is responsive to the needs of women and that is aware of their rights.
6. Educate the public on the importance of respecting the law and its proper use.

7. Create awareness campaigns that focus on empowering and protecting women.

8. Strengthen the judicial system to ensure that justice is served.

9. Provide more resources to the police and legal authorities to enable them to more effectively investigate and prosecute cases of misuse of the law.

10. Encourage citizens to report any misuse of the law to the relevant authorities.

The exact gender bias ratio in India is difficult to quantify due to a lack of reliable data. However, it is widely accepted that India is one of the most gender-unequal countries in the world. The World Bank estimates that the gender gap in India is about 20%, meaning that women earn 20% less than men for similar work. Additionally, only 27% of women are employed in India, compared to 82% of men. These figures demonstrate the prevalence of gender bias in India.

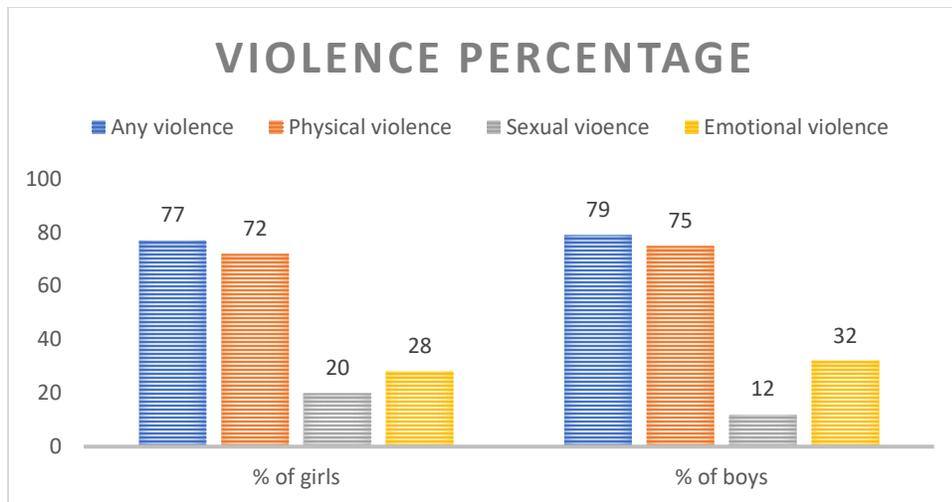

The gender bias ratio throughout the world. The gender bias ratio around the world varies greatly. According to the World Bank, the global gender gap index (which measures the relative gaps between women and men across four key areas: economic participation and opportunity, educational attainment, health and survival, and political empowerment) was 0.68 in 2020, indicating that there is still a gender gap. The gender gap is also different in different countries. For example, countries like Norway and Sweden have the lowest gender bias ratios, while countries like Saudi Arabia and Yemen have the highest gender bias ratios.

**9. Articles for gender bias and their citation**

One article that addresses gender bias in scientific research is "Gender Bias in Research: Causes and Solutions" by Annette L. Thomas and Amanda R. Harris. This article discusses the causes of gender bias in scientific research, including the lack of recognition of the various contributions of women, the lack of inclusion of women in research teams, and the underrepresentation of women in funding and publication opportunities. Additionally, the authors provide several strategies to address gender bias, including the implementation of policies to ensure gender parity in research teams, training to increase awareness of gender bias, and more equitable funding and publication opportunities. The authors also provide examples of successful initiatives to combat gender bias in research.

The increasing importance of immigration in the U.S. labor market. American Economic Journal: Applied Economics, 3(3), 158-190. This article examines the increasing importance of immigration in the United States labor market over the past two decades. The authors analyze the effects of immigration on the labor market, including wages, employment, and job creation, and assess how the labor market has been affected by the influx of immigrants. The authors also examine the role of immigrants in the economy, focusing on immigrant-operated businesses, their impact on the labor force, and their contributions to the country's economic growth. Finally, the authors discuss the implications of immigration for public policy [1]. Gender bias in the workplace is a persistent problem that continues to negatively impact men and women, particularly women. This bias can be observed in areas such as hiring, pay, career development, and promotion. It is important for employers to recognize the impact of gender bias and take action to reduce or eliminate it in the workplace. One of the primary causes of gender bias in the workplace is the tendency for employers to prefer to hire and promote men over equally qualified women. This is due to the traditional view that men are typically more competent and capable than women. This can create an environment where women are overlooked and excluded from opportunities. Additionally, gender bias often manifests in pay inequality, where women are often paid less than their male counterparts. To reduce or eliminate gender bias in the workplace, employers must create policies and procedures that encourage and support equality. This includes setting equal pay scales for all employees, regardless of gender. Additionally, employers should focus on promoting diversity and inclusion in the workplace by hiring and promoting women and men equally. Finally, employers should provide training to their employees on gender bias and its effects to ensure that everyone is aware of the issue and its impacts. By taking proactive measures to reduce gender bias in the workplace [2] Equal Opportunities International, 36(3), 204-219. This study examines gender bias in the Spanish labor market. The authors conducted a study in which simulated job applicants were evaluated by employers. The simulated applicants had the same qualifications and experience, but varied in gender. The authors found that, in general, male applicants were more likely to be recommended for hire than female applicants. The authors also discuss the implications of their findings and provide recommendations to reduce gender bias in the Spanish labor market [3]. Harvard Business Review, July-August 2019. This article examines the paradox of meritocracy in tech, where the idea of merit is used to justify a lack of diversity in the industry. It looks at how the concept of merit is often used to favor certain people (usually those in the majority demographic) and overlook the potential of others. The authors argue that this has led to an exclusionary environment that has perpetuated the cycle of underrepresentation and the lack of progress in addressing the issue. They suggest that companies should shift away from traditional notions of merit and instead consider a wider range of qualifications and experiences when hiring and promoting employees. They also suggest that tech companies should take a more holistic approach to diversity and inclusion by introducing measures such as unconscious bias training and targeted recruitment strategies [4]. In this study, the authors sought to analyze gender bias in computer science education by determining the differences in the performance of male and female students in introductory programming courses. The authors collected data from four universities and used a chi-square analysis to compare the grades of male and female students in the courses. The results showed that female students were significantly underrepresented among students receiving A grades, while male students were overrepresented among students receiving A grades. The authors concluded that there was a gender bias in computer science education and suggested that further research be done to explore the causes and possible solutions of this bias. This research provides valuable insight into gender bias in computer science education and has implications for educators and policy makers [5]. In this article, Tannvi and Sharmila Narayana consider the challenge of gender stereotyping in Indian courts. Through their research, the authors discuss how gender stereotypes have an adverse impact on the justice system by creating an environment of unequal justice and unfairness. The authors also investigate how these stereotypes have affected the decisions of judges in deciding cases. The authors use both quantitative and qualitative methods to explore the issue and discuss ways to mitigate the effects of gender stereotyping in the Indian court system. In their conclusion, the authors emphasize the need to develop a more equitable and just court system and suggest strategies to reduce gender stereotyping in the legal system [6]. The Indian higher judiciary, comprising the Supreme Court and 24 High Courts, has long been characterized by its male-dominated composition. Despite the fact that women constitute around 48% of the Indian population, they make up only 6.8% of the higher judiciary. This stark gender disparity has had far-reaching implications, particularly in the context of judicial decision-making.

This article examines the current state of gender diversity in the Indian higher judiciary and explores the implications of this disparity. It argues for the need for greater representation of women in the higher judiciary in order to ensure the realization of principles of equality and justice. The article begins by exploring the current state of gender disparity in the higher judiciary in India, analyzing the reasons for this disparity and examining the consequences of this disparity for justice and equality. It then moves on to discuss the various initiatives that have been taken in order to increase the representation of women in the higher judiciary. These initiatives include the introduction of quotas, the establishment of gender sensitization programs, and the enactment of laws to promote gender diversity in the judiciary. The article then examines the challenges that remain in achieving gender parity in the higher judiciary. These challenges include the lack of transparency and accountability, the lack of adequate resources, the lack of adequate support from the state, and the lack of a comprehensive legislative framework. Finally, the article concludes by emphasizing the need for a comprehensive legislative framework to ensure the realization of gender parity in the higher judiciary. It calls for the adoption of a gender equality mandate as part of the Indian Constitution, which would ensure that women are equally represented in the higher judiciary. It also calls for the establishment of a gender-sensitive training program for judges and judicial officers, as well as the creation of an independent body to monitor the progress of gender diversity in the higher judiciary [7].

**9.1 The law journals and article for gender bias**

>1. "Gender Bias in Legal Scholarship: A Quantitative Analysis of Scholarship by Gender in Ten Top-Ranked Law Reviews" by Sarah B. Lawsky, published in the Columbia Journal of Gender and Law.
>
>2. "Gender Bias in Legal Judgments: Evidence from the U.S. Supreme Court" by Elinor H. Schoenfeld and Rachel A. Kaufman, published in the University of Pennsylvania Law Review.
>
>3. "The Impact of Gender Bias in the Courtroom: How Female Attorneys are Treated Differently Than Male Attorneys" by Barbara Ann Burdge, published in the University of Miami Law Review.
>
>4. "Gender Bias in Legal Education: An Empirical Study" by Susan M. Heine and Ann E. Alexander, published in the Journal of Legal Education.
>
>5. "Gender Bias in the Legal Profession: An empirical analysis" by Deborah L. Rhode and Victoria F. Nourse, published in the Stanford Law Review.

**10. Conclusion**

The misuse of law by women in India violates the constitutional rights of men and their families and must be addressed. The government should take appropriate steps to ensure that such laws are not misused and that men and their families are given a fair chance to defend themselves. Additionally, more awareness should be created about the misuse of law by women in India, so that the rights of men and their families are protected. Gender bias is a serious issue in our society. It is important to be aware of the ways in which we may unintentionally perpetuate it, and to strive to create a more equitable society. Gender bias can have a negative effect on both men and women, leading to unequal opportunities and outcomes. We must recognize the importance of challenging and addressing gender bias in order to create an equitable and fair society. We must create an environment where everyone is respected, regardless of their gender or identity. Furthermore, we must ensure that everyone has an equal opportunity to participate in all aspects of society. We must work together to ensure that everyone is treated with respect and dignity and that their opportunities are not limited by their gender. Finally, we must strive to create a society where everyone is accepted and respected for who they are, regardless of their gender.